\documentclass[a4paper,twoside]{article}
\usepackage[utf8]{inputenc}
\usepackage[T1]{fontenc} 
\usepackage{RR,RRthemes}
\usepackage[dvips, 
  a4paper,
  hypertexnames=true, 
  backref=section, 
  colorlinks=true, 
  citecolor=blue,
  bookmarks=true, 
  bookmarksopen=true, 
  pdftitle=Adding\ Network\ Coding\ Capabilities\ to\ the\ WSNet\ Simulator, 
  pdfauthor=Wei\ Liang\ Choo\ -\ Frédéric\ Le\ Mouël\ -\ Katia\ Jaffrès-Runser\ -\ Marco Fiore, 
  pdfsubject=INRIA\ Technical\ Report, 
  pdfcreator=dvipdf,
  pdfkeywords=Network\ Coding\ -\ Wireless\ Sensor\ Network\ (WSN)\ -\ Simulation\ -\ WSNet]{hyperref}
\usepackage{graphicx}
\usepackage{listings}

\RRdate{March 2011}

\RRauthor{
Wei Liang Choo\thanks[univ]{University of Lyon, INSA-Lyon, CITI, F-69621, Villeurbanne, France, \texttt{\{firstname.lastname\}@insa-lyon.fr}}
\and
Frédéric Le Mouël\thanksref{univ}
\thanks[inria-amazones]{INRIA, AMAZONES Team}
\and
Katia Jaffrès-Runser\thanksref{univ}
\thanks[inria-swing]{INRIA, SWING Team}
\and
Marco Fiore\thanksref{univ}
\thanksref{inria-swing}
}

\authorhead{Choo \& Le Mouël \& Jaffrès-Runser \& Fiore}

\RRtitle{Implantation d'un module de codage de réseaux dans le simulateur WSNet}

\RRetitle{Adding Network Coding Capabilities \\ to the WSNet Simulator}

\titlehead{Adding Network Coding Capabilities to the WSNet Simulator}

\RRresume{Ce rapport technique présente l'implantation d'un module de codage de réseaux dans le simulateur de capteurs sans-fil WSNet. L'objectif de conception de ce module est d'avoir une interface de programmation générique pour ensuite spécialiser facilement différentes stratégies de codage: aléatoire, orienté-source/destination, intra/inter-flux, etc.}

\RRabstract{This technical report presents the implementation of a Network Coding module in WSNet - a Wireless Sensor Network simulator. This implementation provides a generic programming interface to allow an easy specialization of different coding strategies: random, source/destination-oriented, intra/inter-flow, etc.}

\RRmotcle{Codage des réseaux, Réseaux de capteurs sans-fil, Simulation, WSNet}

\RRkeyword{Network Coding, Wireless Sensor Network (WSN), Simulation, WSNet}

\RRthemeProj{swing}
\RRprojet{SWING}

\URRhoneAlpes 

\RRNo{0405}

\begin{document}

\makeRT 

\tableofcontents
\newpage

\section{Introduction}

Our goal is to study the impact of different Network Coding strategies~(NC) on end-to-end service delivery over mobile and wireless Disruption-Tolerant Networks~(DTNs). To realize this study, in a first step, we simulate a mobile and wireless DTN environment. This report presents (i) in section~\ref{section:context}, the context: DTN, NC and why we have chosen the WSNet simulator, (ii) in section~\ref{section:nc-module}, our NC framework provided in a WSNet module: architecture, generic API definition, packet storing, linear independence checking, real encoding/decoding.

\section{Context}
\label{section:context}

\subsection{Disruption-Tolerant Network}

Disruption-Tolerant Networking~(DTN) is an approach that seeks to address the non-constant nature of links in networks~\cite{Fall:2003:DNA:863955.863960}. This could be caused by mobility of the nodes or interference in the environment. 

\begin{figure}[!hbt]
\centering
\includegraphics[width=0.75\textwidth]{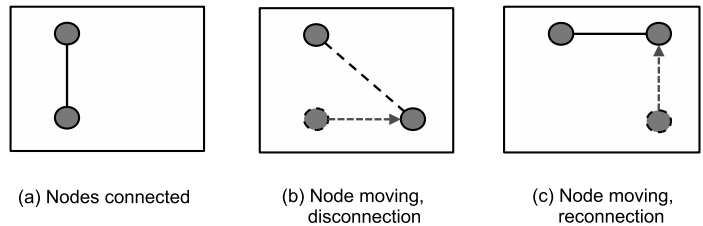}
\caption{Disrupted Network Example}
\label{fig:dtn}
\end{figure}

Figure~\ref{fig:dtn} shows an example of DTN where, as a node moves, the link previously established is broken. The link may be established again once the node moves back into range of communicating with the original node. Conventional routing involves finding a path and forwarding packets from the source to the destination. However because of the link break, storing and then forwarding when the link is re-established may be needed - so introducing a delay. A common approach is to send out replicated packets to many nodes hoping that packets will reach the destination. However, this takes up large amounts of storage and bandwidth.

Strategies involving Network Coding are to be accessed to judge their impacts on delay/tolerance/capacity improvement of a DTN environment~\cite{Zhang06, 4509805, Altman:2010:DCC:1833515.1833540}.

\subsection{Network Coding}

Network Coding~(NC) is a technique where nodes of a network are able to combine together two more or received packets and transmit them~\cite{Fragouli:2006:NCI:1111322.1111337}. With enough information - enough encoded packets, the original packets can then be decoded at the destination. This is a change from just forwarding packets which can bring about potential throughput improvements and a high degree of robustness.

\begin{figure}[!hbt]
\centering
\includegraphics[width=0.65\textwidth]{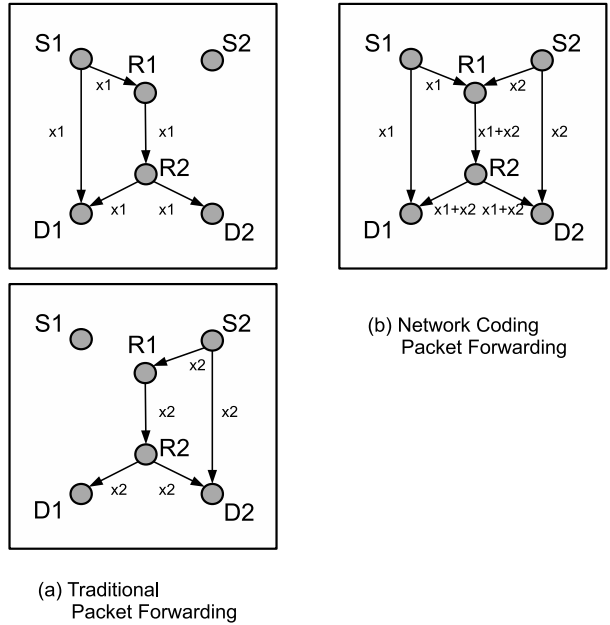}
\caption{Network Coding Butterfly Example}
\label{fig:nc}
\end{figure}

Figure~\ref{fig:nc} presents the Butterfly Network Coding multicast example. S1 and S2 are Source nodes, R1 and R2 are Relay nodes, D1 and D2 are Destination nodes and x1 and x2 are packets from S1 and S2 respectively. D1 and D2 needs to receive both x1 and x2 packets. In the traditional packet sending method, x1 would be sent to D1 from S1 directly. But due to distance from D2, S1 will send x1 via R1 to R2 and then to D2. The same case is for S2 and D1. R1 forwards the whole packet x1 then x2, to R2 which then forwards it to D1 and D2. 
Using NC, when R1 receives both x1 and x2, R1 can combine the packets and send only one packet combining both x1 and x2 to R2 which forwards it to D1 and D2. D1 and D2 use this encoded packet to retrieve the other missing packet. In this case, NC thus helps in reducing the sending of a second packet from R1 and R2.

According to the network topologies considered (linear vs non-linear, multicast vs non-multicast, directed vs undirected, cyclic vs acyclic), different NC strategies exist to select and encode packets~\cite{katti2005}: random, unique/multi source-oriented, unique/multi destination-oriented, intra-session, inter-session, etc. We plan to develop and test social and service-oriented NC strategies and so we need a realistic simulation environment to compare them in a mobile and wireless DTN.

\subsection{WSNet Simulator}

WSNet is a simulator for large-scale Wireless Sensor Networks created and developed at the CITI Laboratory~\cite{Fraboulet:2007:WDP:1236360.1236385}. While several simulator exist for DTN~\cite{RomeroAmondaray:2008:DTN:1409985.1410006, keranen-theone}, WSNet main features - Node Simulation, Environment Simulation, Radio Medium Simulation and Extensibility - are particularly suitable for our NC testing. Radio medium simulation provides realistic radio channel modeling appropriate to test wireless communication in mobile DTN. Node Simulation allows the integration of the application level, suitable to test social and service-oriented NC strategies.

\begin{figure}[!hbt]
\centering
\includegraphics[width=0.55\textwidth]{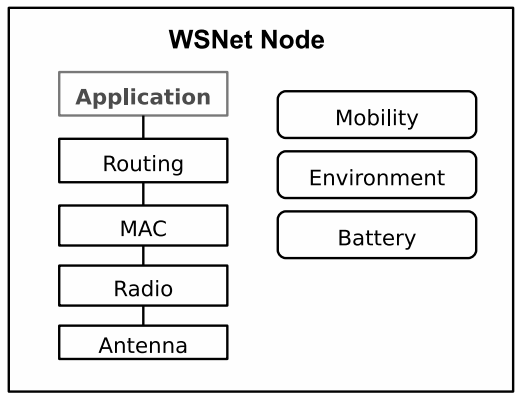}
\caption{Modular Architecture of a WSNet Node}
\label{fig:wsnet}
\end{figure}

Figure~\ref{fig:wsnet} describes a node architecture that can be created in WSNet. There are already various standing modules that can be used for each part: support for complex nodes architecture~(MIMO systems, multiple radio/antenna interface support), support for energy consumption simulation, support for various propagation models, support for propagation delays, etc. Modules are attached on run time and an XML file is used to control the WSNet. 

\section{A Generic Network Coding Module in WSNet}
\label{section:nc-module}

\subsection{Module Configuration}

Using the WSNet extensibility feature~\cite{BenHamida2007a}, we have developed an application module that simulates a wireless DTN with NC by storing, selecting/dropping, encoding/decoding IP packets.

\begin{figure}[!hbt]
\centering
\includegraphics[width=0.55\textwidth]{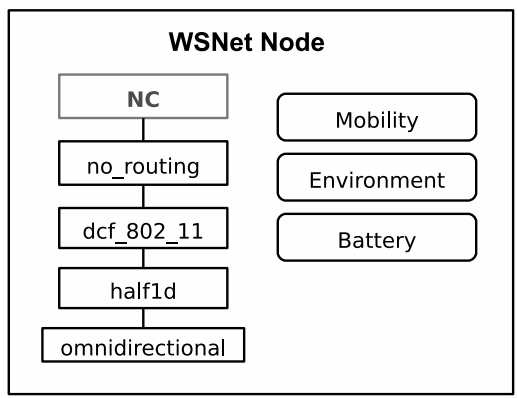}
\caption{Network-Coding Module in WSNet}
\label{fig:wsnet-nc}
\end{figure}

We configure and test it with different existing WSNet modules~\cite{BenHamida2007b} (cf Figure~\ref{fig:wsnet-nc} and Listing~\ref{list:confxml}). No routing module has currently been used since a static one, with the Butterfly example, was applied.

\begin{lstlisting}[language=XML,caption=DTNNC\_Butterfly\_example.xml,label=list:confxml]
<!-- == RADIO and ANTENNA ============================== -->
<entity name="omnidirectionnal" library="antenna_omnidirectionnal">
  <default loss="0" angle-xy="random" angle-z="random"/>
</entity>

<entity name="radio" library="radio_half1d">
  <default sensibility="-92" T_b="727" dBm="10" channel="0" modulation="none"/>
</entity>

<!-- == MAC ============================================ -->
<entity name="mac" library="mac_dcf_802_11">
</entity>

<!-- == APPLICATION ==================================== -->
<entity name="Esource0" library="application_DTNNC_Source">
  <init debugMode="1" dataStruct="1" maxNumberOfCombinedPerPacket="30" maxNumberOfPackets="100" numberOfCoefficients="2" FPower="3" storageOrder="1" decodingPolicy="1" encodingType="1" encodingPacSelection="1" chanceOfSending="10"/>
  <default type="0" nodenum="0" period="2s" inData="15"/>
</entity>

<entity name="Esource1" library="application_DTNNC_Source">
  <init debugMode="1" dataStruct="1" maxNumberOfCombinedPerPacket="30" maxNumberOfPackets="100" numberOfCoefficients="2" FPower="3" storageOrder="1" decodingPolicy="1" encodingType="1" encodingPacSelection="1" chanceOfSending="10"/>
  <default type="0" nodenum="1" period="2s" inData="15"/>
</entity>

<entity name="Esensor0" library="application_DTNNC_Sensor">
  <init debugMode="1" dataStruct="1" maxNumberOfCombinedPerPacket="30" maxNumberOfPackets="100" numberOfCoefficients="2" FPower="3" storageOrder="1" decodingPolicy="1" encodingType="1" encodingPacSelection="1" chanceOfSending="100"/>
  <default type="1" nodenum="0" period="2s" inData="15"/>
</entity>

<entity name="ErelayDumb0" library="application_DTNNC_RelayDumb">
  <init debugMode="1" dataStruct="1" maxNumberOfCombinedPerPacket="30" maxNumberOfPackets="100" numberOfCoefficients="2" FPower="3" storageOrder="1" decodingPolicy="1" encodingType="1" encodingPacSelection="1" chanceOfSending="100"/>
  <default type="3" nodenum="0"/>
</entity>

<entity name="Esink0" library="application_DTNNC_Sink">
  <init debugMode="1" dataStruct="1" maxNumberOfCombinedPerPacket="30" maxNumberOfPackets="100" numberOfCoefficients="2" FPower="3" storageOrder="1" decodingPolicy="1" encodingType="1" encodingPacSelection="1" chanceOfSending="0"/>
  <default type="2" nodenum="0"/>
</entity>

<entity name="Esink1" library="application_DTNNC_Sink">
  <init debugMode="1" dataStruct="1" maxNumberOfCombinedPerPacket="30" maxNumberOfPackets="100" numberOfCoefficients="2" FPower="3" storageOrder="1" decodingPolicy="1" encodingType="1" encodingPacSelection="1" chanceOfSending="0"/>
  <default type="2" nodenum="1"/>
</entity>

\end{lstlisting}

\subsection{Architecture Overview}

\begin{figure}[!hbt]
\centering
\includegraphics[width=\textwidth]{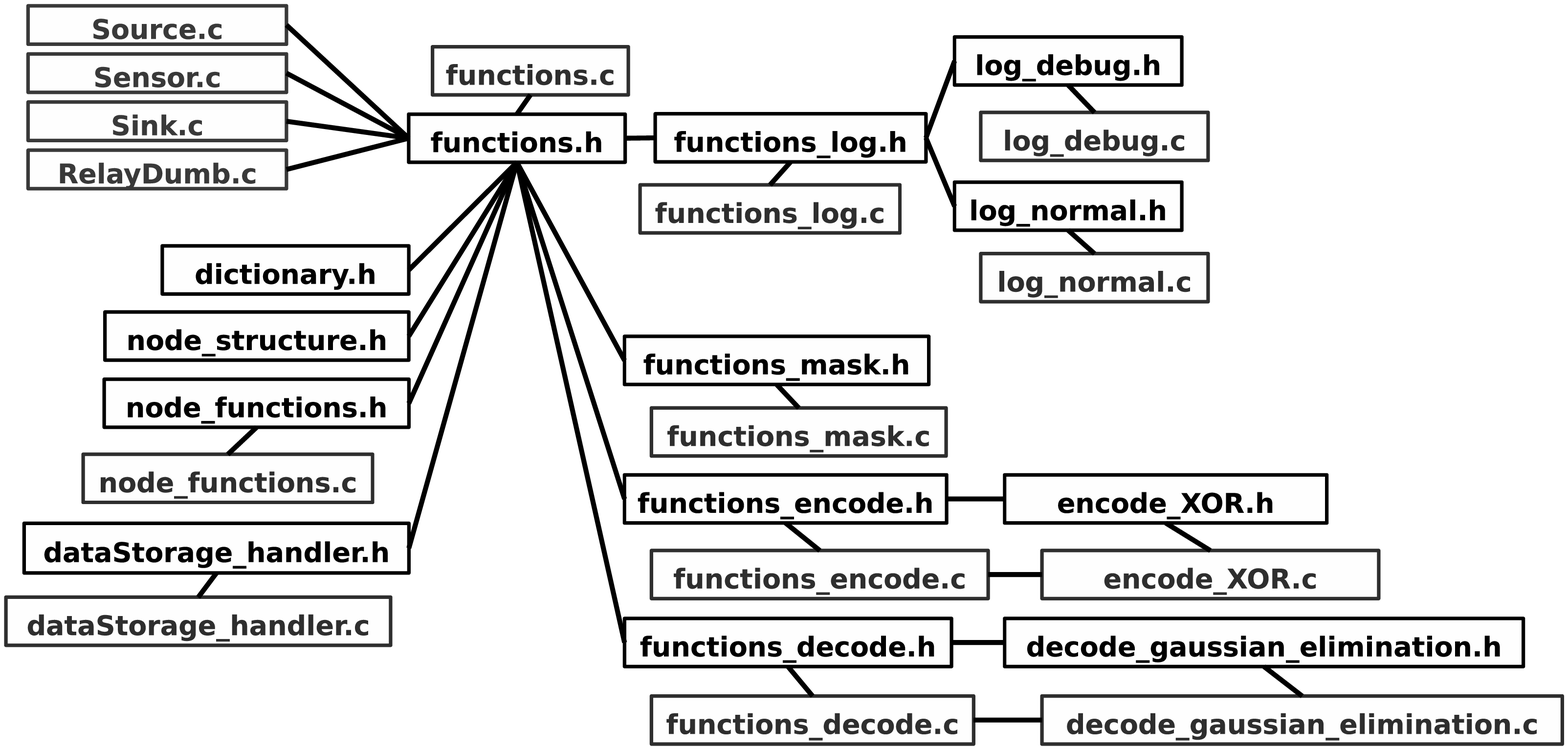}
\caption{Network-Coding Module Architecture}
\label{fig:wsnet-nc-architecture}
\end{figure}

The Figure~\ref{fig:wsnet-nc-architecture} presents the architecture of the NC module. Each node includes the \texttt{functions.h} header file, entry point of the framework. This file defines (i) the common node structure, variables and data storage - detailed in section~\ref{section:node}, (ii) the common masking/encoding/decoding functions - detailed in section~\ref{section:nc-api} and (iii) useful logging functions - detailed in section~\ref{section:log-api}.

\subsection{Node Definition}
\label{section:node}

\subsubsection{Node Type}

All common aspects of a node are included in \texttt{DTNNC\_dictionary.h}, \texttt{DTNNC\_node\_structure.h}, \texttt{DTNNC\_functions.h}, \texttt{DTNNC\_functions.c} and \texttt{DTNNC\_dataStorage\_handler.h}. Each node can then be specialized and instantiated to be a:

\begin{description}
\item[\emph{Source}] \hfill
  \begin{itemize}
  \item Creates packets
  \item Sends packets
  \end{itemize}
\newpage
\item[\emph{Sensor}] \hfill
  \begin{itemize}
  \item Stores received packets
  \item Encodes packets, including masking and xor-ing packets
  \item Decodes encoded packets, if possible and needed
  \item Sends encoded packets out
  \end{itemize}
\item[\emph{RelayDumb}] \hfill
  \begin{itemize}
  \item Forwards received packets
  \end{itemize}
\item[\emph{Sink}] \hfill
  \begin{itemize}
  \item Stores received packets
  \item Decodes 
  \end{itemize}
\end{description}

\subsubsection{Node Common Variable Definitions: \texttt{dictionary.h}}

\texttt{DTNNC\_dictionary.h} defines keywords to make code more readable and easier to use.

\begin{lstlisting}[language=C,caption=\texttt{DTNNC\_dictionary.h},label=list:ndict]
#ifndef __DTNNC_dictionary__
#define __DTNNC_dictionary__

/* Node Mode */
#define SOURCE 0
#define SENSOR 1
#define SINK 2
#define RELAYDUMB 3

/* Log Mode */
// change to 1 for debug
// change to 0 for normal
#define DEBUG_MODE 1

/* Debug output prints */
#define DEBUG_MASK 0
#define DEBUG_DECODING 1
#define DEBUG_NODE_INFO 1

/* Debug LogFunction */
#define PRINT_NODE_STORAGE 40
#define PRINT_DATA_HOLDER 41
#define PRINT_RECONSTRUCTED_PACKETS 42

/* LogFunction */
#define PRINT_SEND_EVENT 1
#define PRINT_RECEIVE_EVENT 2
#define PRINT_STORE_PACKET 3
#define PRINT_DROP_PACKET_REPEAT 4
#define PRINT_DROP_PACKET_DECODED 5
#define PRINT_LINEAR_CHECK_FAILED 6
#define PRINT_LINEAR_CHECK_PASSED 7
#define PRINT_DECODED 8
#define PRINT_ENCODED 9
#define PRINT_UNSETNODE 10 // to be changed to more detailed

#endif // __DTNNC_dictionary__
\end{lstlisting}

\subsubsection{Node Structure: \texttt{node\_structure.h}}

\texttt{DTNNC\_node\_structure.h} defines the common structure of all nodes. It is easier to edit one structure for all nodes than to make a specific structure for each node type. Most variables are needed in all node types.

\begin{lstlisting}[language=C,caption=\texttt{DTNNC\_node\_structure.h},label=list:nstruc]
#ifndef __DTNNC_node_structure__
#define __DTNNC_node_structure__

/* Node private data */
struct nodedata {

  int *overhead;
  // Source, Sensor, Sink or RelayDumb
  int type; 
  // ID of the node. i.e. node 0 of Source type
  int nodenum; 
  // Flag if Node has decoded packets
  int decodedFlag; 

  int seqNum;

  // period in which to recall node
  uint64_t period; 

  /* temp data holder */
  struct packet_data* dataHolder;
  struct header_packets_combined ** dataHolder_header_packets_combined;
  uint16_t dataHolder_numberOfpacketsCombined;

  int* dataHolder_args;

  /* data storage */
  // data storage type 1(basic arrays)
  struct packet_data** stored_data; 
  struct header_packets_combined *** stored_data_header_packets_combined;
  uint16_t* stored_data_numberOfpacketsCombined;

  /* arguments storage */
  int** stored_data_args; 

  /* reconstructed packets storage */
  struct packet_data** reconstructed_pack_data;
  int** reconstructed_pack_args;

  /* num of packets stored counter */ 
  int num_of_packets_stored;

  /* for stats */
  // Number of packets transmited by node
  int packet_tx;
  // Number of packets received by node
  int packet_rx;

};

#endif // __DTNNC_node_structure__
\end{lstlisting}

\subsubsection{Node Management Functions: \texttt{node\_functions.h}}

\texttt{DTNNC\_node\_functions.h} allows to manage a node. It creates the variables, allocates the memory for variable structures, and sets the variables to default. This API simplifies the development. Editing or adding a new node variable requires the editing and adding of this variable in each node type’s source code file. With this API, it requires to be done only one time in one place.

\begin{lstlisting}[language=C,caption=\texttt{DTNNC\_node\_functions.h},label=list:nfunc]
#ifndef __DTNNC_node_functions__
#define __DTNNC_node_functions__

#include "DTNNC_functions.h"

// Called when setting Node Entity (from int())
int setNodeEntity(call_t *c, void *params);

// Called when setting Node Variables (from setnode())
int setNodeVariables(call_t *c, void *params);

// Called when unsetting node(from unsetnode())
// frees memory allocated
int freeNode(call_t *c);

#endif // __DTNNC_node_functions__
\end{lstlisting}

\subsubsection{Node Storage: \texttt{packet.h} \texttt{dataStorage\_handler.h}}

Data stored are IP packets. These packets are potentially xor-mixed packets, so a packet header includes the number and a table of sub-packet headers. A final sub-packet header contains an id (can be an application or a service id), a sequence number ordering a packet flow, a source and $n$ destinations (for broadcast or multicast). Packet data structure contains the real data (here a dummy example with 4 characters).

\begin{lstlisting}[language=C,caption=\texttt{DTNNC\_node\_dummy\_packet.h},label=list:npacket]
#ifndef __DTNNC_node_dummy_packet__
#define __DTNNC_node_dummy_packet__

/* Packet header */
struct packet_header {
	struct header_packets_combined ** header_packets_combined_info;
	uint16_t numberOfpacketsCombined;
	uint16_t *args;
};

struct header_packets_combined {
	uint16_t seqNum;
	uint16_t id;
	uint16_t source;
	uint16_t numOfDest;
	uint16_t *destination;
};

/* Dummy packet data*/
struct packet_data{
	unsigned char packetdata;
	unsigned char packetdataB;
	unsigned char packetdataC;
	unsigned char packetdataD;
};

#endif // __DTNNC_node_dummy_packet__
\end{lstlisting}

\texttt{DTNNC\_dataStorage\_handler.h} defines generic functions providing the storage functionality. The data storage structure can easily be changed without changing many other code source files. Accessing data is allowed by using get and set methods and not accessing to the data directly.

\begin{lstlisting}[language=C,caption=\texttt{DTNNC\_dataStorage\_handler.h},label=list:nstorage]
#ifndef __DTNNC_dataStorage_handler__
#define __DTNNC_dataStorage_handler__

#include "DTNNC_functions.h"

// dataHolder get and set methods
void* getDataHolderDataAt(call_t* c);

void* getDataHolderData_packetInfoAt(call_t*c, int positionInHeader);

int getDataHolder_numberOfpacketsCombinedAt(call_t*c);

int setDataHolderDataAt(call_t* c, struct packet_data* dataT);

int setDataHolderData_packetInfoAt(call_t*c, int positionInHeader, struct header_packets_combined* tempHPC);

int setDataHolder_numberOfpacketsCombinedAt(call_t*c, int numberOfpacketsCombinedIn);

// dataStorage get and set methods

// Get method to get packet data
void* getDataAt(call_t* c, int position);

void* getData_packetInfoAt(call_t*c, int positionInDataStorage, int positionInHeader);

int getData_numberOfpacketsCombinedAt(call_t*c, int position);

// Set method to set packet data
int setDataAt(call_t* c, int position, struct packet_data* dataT);

int setData_packetInfoAt(call_t*c,int positionInDataStorage, int positionInHeader, struct header_packets_combined* tempHPC);

int setData_numberOfpacketsCombinedAt(call_t*c, int position, int numberOfpacketsCombinedIn);

// data Handling functions

// Generic method to add packet received to storage
// Check if repeat packet is already stored.
// Defination of non repeat packet is received packet 
// contains an int not found in that column of storage
// Check storage strategy
int addAllData(call_t * c, packet_t * packet);

// use set methods to add data.
// for adding in FIFO other.
int addAllDataFIFO(call_t *c, packet_t *packet);

// Swap data between 2 specified data in storage
int swapRowAllData(call_t * c, int dataA, int dataB);

// Reconstructed packet storage get and set methods

// Get method to get data in reconstructed storage
void* getRCDataAt(call_t* c, int position);

// Set method to set data in reconstructed storage
int setRCDataAt(call_t* c, int position, struct packet_data* dataT);

#endif // __DTNNC_dataStorage_handler__
\end{lstlisting}

\subsection{Network-Coding API Definition}
\label{section:nc-api}

\subsubsection{Node NC Entry Point: \texttt{functions.h}}

\texttt{DTNNC\_functions.h} header file is the main entry point of our framework and connects all other files of the module. It defines the four functionalities of one node: storing / dropping / encoding / decoding IP packets. \\

\noindent Creation of a node only needs to include \texttt{DTNNC\_functions.h}

\begin{lstlisting}[language=C,caption=\texttt{DTNNC\_functions.h},label=list:func]
#ifndef __DTNNC_functions__
#define __DTNNC_functions__

#include <stdio.h>
#include <include/modelutils.h>

#include "DTNNC_dictionary.h"
#include "DTNNC_node_structure.h"
#include "DTNNC_node_entity.h"
#include "DTNNC_node_dummy_packet.h"

// The headers files below have included this header file
//#include "DTNNC_dataStorage_handler.h"
//#include "DTNNC_functions_encode.h"
//#include "DTNNC_functions_decode.h"
//#include "DTNNC_functions_log.h"
//#include "DTNNC_functions_mask.h"

/* Functions */

/* Encodes packets */
// Calls encodeFunction() from "DTN_functions_encode.h"
// which does network coding on two packets depending on policy set
// Stores resultant packet into dataHolder storage
int encode(call_t* c);

/* Decodes packets */
// Calls decodeFunction() from "DTN_functions_decode.h"
// which attempts to do the decoding
// stores decoded packets into reconstructed packets storage
int decode(call_t* c);

/* Stores received packet */
// Checks if node has decoded packets. if yes than drop received packet
// Copy received packet to dataHolder storage, This is for sending(read report)
// Calls addAllData() from "DTN_dataStorage_handler.h" 
// which stores the packet data structure
int store(call_t* c, packet_t* packet);

/* Drops packets */
// Have not been used
// Currently empty
int drop(call_t* c);

#endif // __DTNNC_functions__
\end{lstlisting}

\subsubsection{NC Masking: \texttt{functions\_mask.h}}

Before combining different packets, a choice of fragmenting data information of one packet can be done. Part of the data can be kept and part can be 'masked'. We use the word 'mask' for the randomly chosen coefficient used in the random linear network coding ($p = \sum_i \lambda_i p_i$, with $p_i$ the packet fragments, $\lambda_i$ the coefficients which are referred to in the following as 'masks'). \texttt{DTNNC\_functions\_mask.h} offers this masking function.

\begin{lstlisting}[language=C,caption=\texttt{DTNNC\_functions\_mask.h},label=list:funcm]
#ifndef __DTNNC_functions_mask__
#define __DTNNC_functions_mask__

#include "DTNNC_functions.h"

/* Mask data */
// mask data chosen from data structure (dataChoice)
// which holds the column in the matrix (coefficientCol)
// with the mask (maskA)
// input maskA of 0 means a random mask
int mask(call_t *c,int dataChoice,int coefficientCol,int maskA);

#endif // __DTNNC_functions_mask__
\end{lstlisting}

\subsubsection{NC Masking - one implementation: \texttt{functions\_mask.c}}

\begin{figure}[!hbt]
\centering
\includegraphics[width=\textwidth]{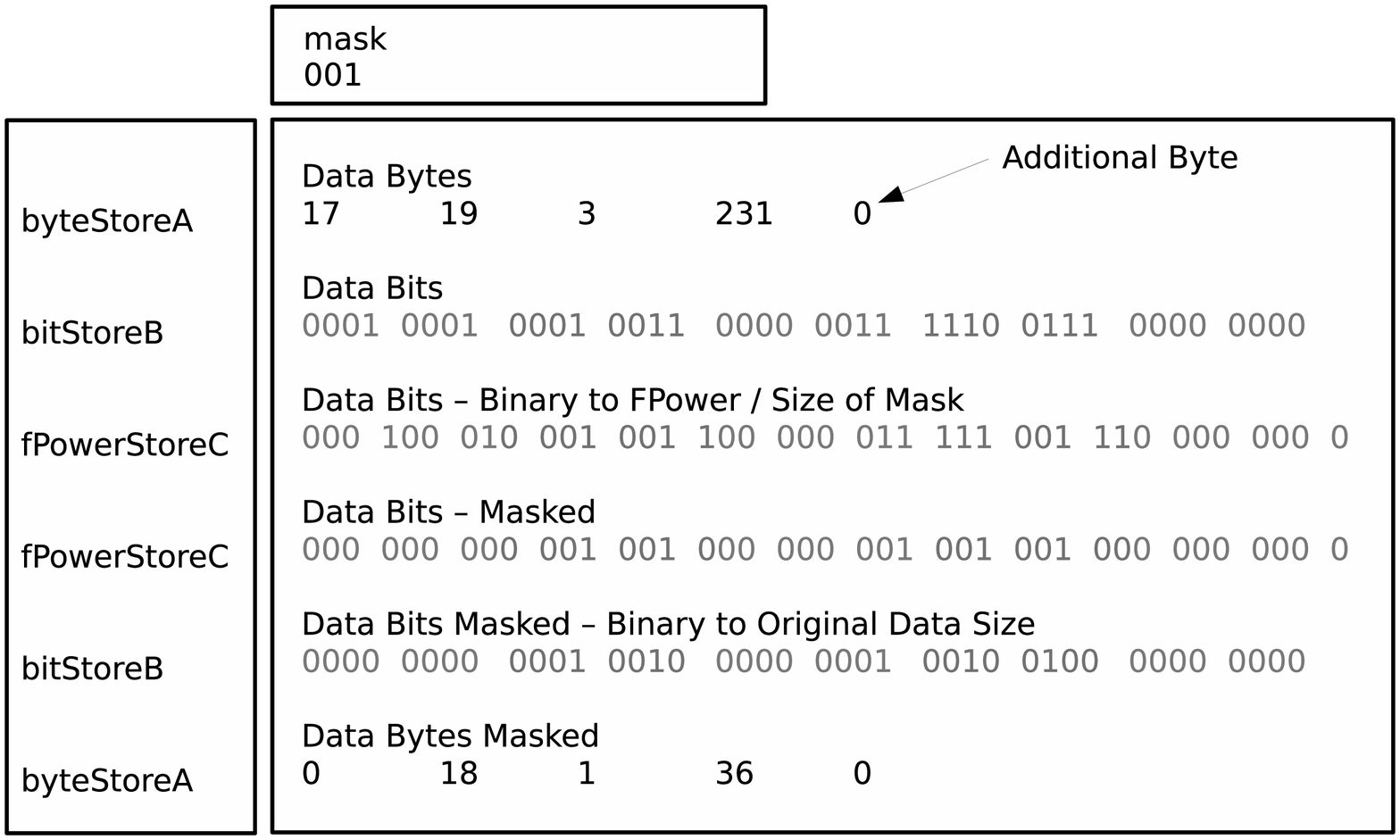}
\caption{One Data Masking Implementation: One Additional Byte / Byte to Bit conversion / Flow masking}
\label{fig:masking}
\end{figure}

There are various methods that can be used to mask packet data. Upon consideration, the method described in Figure~\ref{fig:masking} is used.

One additional byte is added to the data packet during the packet memory allocation. The purpose is to tackle the problem of improper masking when the total bit size of the packet data is not a factor of the mask size. 

Packet data is copied byte by byte into a byte storage (byteStoreA). Data in byteStoreA is then transferred bit by bit into a bit storage (bitStoreB). Data in bitStoreB is then transferred bit by bit into a FPower/Size of Mask storage (fPowerStoreC) (in Figure~\ref{fig:masking}, size of the mask is 3). fPowerStoreC is then masked using the selected mask (in Figure~\ref{fig:masking}, the mask is 001). fPowerStoreC is then transferred back bit by bit to bitStoreB. bitStoreB is transferred bit by bit to byteStoreA. byteStoreA is then copied byte by byte into the dataHolder. Finally, the mask is stored into the argument storage of the dataHolder.

\subsubsection{NC Encoding: \texttt{functions\_encode.h}}

\texttt{DTNNC\_functions\_encode.h} encodes data depending on an encoding strategy.
	
The \texttt{encodeFunction} acts as a controller function. When this function is called, it first checks the encoding validity and then calls the real specific encoding function. Currently the checking consists in testing the node role: if the node is only a relay, it checks if there is only one packet in storage buffer and forwards that packet without changing it; if the node has an encoding role and several packets in the buffer, then the real specific encoding function is called.

\begin{lstlisting}[language=C,caption=\texttt{DTNNC\_functions\_encode.h},label=list:funce]
#ifndef DTNNC_ENCODE_H_
#define DTNNC_ENCODE_H_

#include "DTNNC_functions.h"

// Controller function
// Encoding takes place according to strategy chosen (i.e. random)
// Checks number of packets stored, if only 1 packet in storage than return
// Checks encoding type chosen and calls function for encoding
int encodeFunction(call_t *c);

// Specific function
// Encoding takes place between two specified datas
// This function is used for swapping data, needed in decoding process
// Checks and encoding type chosen and calls function for encoding
int encodeFunctionSpecific(call_t *c, int sourceDataA, int sourceDataB, int destinationData);

#endif /* DTNNC_ENCODE_H_ */
\end{lstlisting}

\subsubsection{NC Encoding - one implementation: \texttt{functions\_encode.c}}

We provide in the module one random XOR encoding implementation in \texttt{DTNNC\_encode\_XOR.h} and \texttt{DTNNC\_encode\_XOR.c}. Two random packets are chosen from the stored data in the node. As \texttt{rand()} of C is biased, an improved version of random is used: seeding of the random number is done at start of node setup at \texttt{init()} of each node type.

\begin{lstlisting}[language=C,caption=\texttt{DTNNC\_encode\_XOR.h},label=list:funcexor]
#ifndef DTNNC_ENCODE_XOR_H_
#define DTNNC_ENCODE_XOR_H_

#include "DTNNC_functions_encode.h"

// Encoding using XOR

// Check encoding strategy and encodes accordingly
// Result stored in dataHolder
int encodeXOR(call_t *c);

// Encode choosing 2 random packets in storage
// Result stored in dataHolder
int encodeXOR_random(call_t *c);

// Encoding using XOR between two specified datas
// Result stored in dataHolder
int encodeXORSpecific(call_t *c, int sourceDataA, int sourceDataB, int destinationData);

#endif /* DTNNC_ENCODE_XOR_H_ */
\end{lstlisting}

\subsubsection{NC Decoding: \texttt{functions\_decode.h}}

\texttt{DTNNC\_functions\_decode.h} decodes data following a decoding strategy.

The \texttt{decodeFunction} acts also as a controller function. It checks linear dependency before decoding: it checks if the last received packet contains enough relevant new information comparing to existing information in the storage buffer. If so the decoding testing is applied.

\begin{lstlisting}[language=C,caption=\texttt{DTNNC\_functions\_decode.h},label=list:funcd]
#ifndef DTNNC_DECODE_H_
#define DTNNC_DECODE_H_

#include "DTNNC_functions.h"

// Controller function
// Decoding takes place according to strategy chosen
int decodeFunction(call_t * c);

#endif /* DTNNC_DECODE_H_ */
\end{lstlisting}

\subsubsection{NC Decoding - one implementation: \texttt{functions\_decode.c}}

We provide in the module one Gaussian Elimination implementation in \texttt{DTNNC\_decode\_gaussian\_elimination.h} and \texttt{DTNNC\_decode\_gaussian\_elimination.c} (cf Listing~\ref{list:funcdgauss}). The \texttt{linearIndependentCheck} function implements the Linear Independence Checking; the \texttt{forwardSubstitution} function implements the first phase of the Gaussian method: the Forward Substitution; the \texttt{reverseElimination} function implements the second phase of the Gaussian method: the Reverse Elimination; the \texttt{reconstructPacket} function finally implements the third phase of the Gaussian method and retrieves the original packet.

\begin{lstlisting}[language=C,caption=\texttt{DTNNC\_decode\_gaussian\_elimination.h},label=list:funcdgauss]
#ifndef DTNNC_DECODE_GAUSSIAN_ELIMINATION_H_
#define DTNNC_DECODE_GAUSSIAN_ELIMINATION_H_

#include "DTNNC_functions_decode.h"

// decoding using the created gaussian elimination method
// 0) linear independence (return 0 and stop decoding if not linearly independent)
// 1) Forward Substitution
// 2) Reverse Elimination
// 3) Reconstruct packet
int decode_gaussian_elimination(call_t *c);

// Forward substitution, step 1 of gaussian elimination
// Changes the argument matrix
// Use linearIndependentCheck() as guard to ensure linear independent
int forwardSubstitution(call_t *);

// Reverse elimination, step 2 of gaussian elimination
int reverseElimination(call_t *);

// Reconstruction of packet, step 3 of gaussian elimination
// Store reconstructed packets in reconstruction data holder
// Node data storage stores the data up to Step2 (data of each bit)
int reconstructPacket(call_t *);

// Check that the argument matrix is linearly independent
// Return 0 if non linearly independent
int linearIndependentCheck(call_t *);

#endif /* DTNNC_DECODE_GAUSSIAN_ELIMINATION_H_ */
\end{lstlisting}

We illustrate the Gaussian implementation with an example in Figures~\ref{fig:fs1}-\ref{fig:rc2}: a masking in the finite group $F_{2^3}$ has been applied on 2 packets from different sources.

\begin{enumerate}
\item Checking Linear Independence

Checking Linear Independence is basically the same as the Forward Substitution process and code. However doing forward substitution corrupts the matrix and if the modified matrix is not linearly independent, the original matrix cannot be retrieved easily.

Therefore the first step of checking linear independence phase is to clone the matrix into a temporary matrix for testing linear independence. As this is just a checking phase, data is not touched and therefore not cloned. Check for linear independence fails when the matrix lines swap is not successful, meaning a triangulation can not be performed. 

\item Forward Substitution

Figures~\ref{fig:fs1} to~\ref{fig:fs3} show how forward substitution is done. It starts from the top of the matrix and works in the binary format (even if the matrix is stored in an integer format).

\begin{itemize} 
\item Swap phase \hfill \\
If the first bit of the first column is not at 1, the algorithm finds the first row containing this 1 and data of the rows are swapped.

\begin{figure}[!hbt]
\centering
\includegraphics[width=0.7\textwidth]{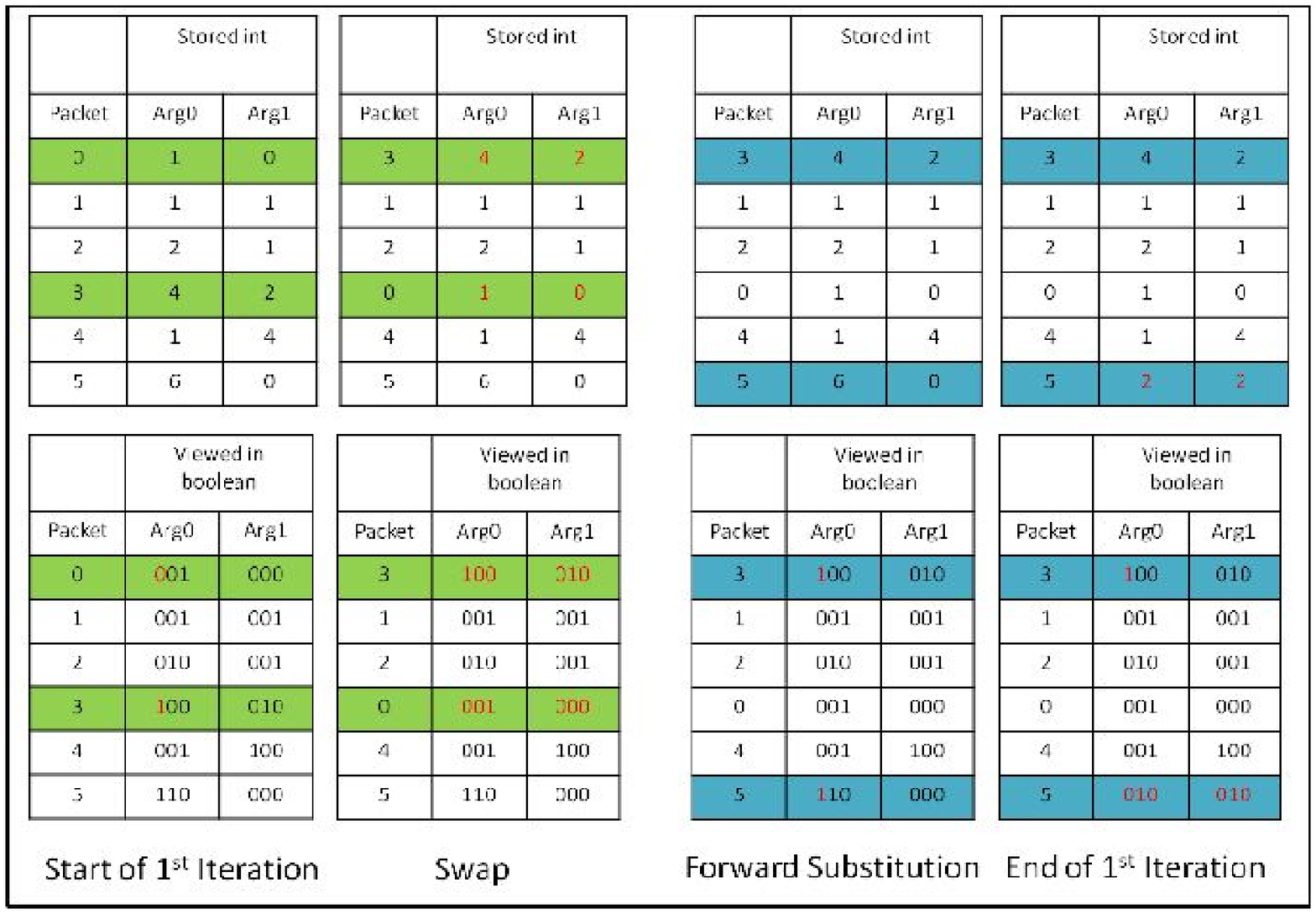}
\caption{First step of Forward Substitution phase}
\label{fig:fs1}
\end{figure}

\begin{figure}[!hbt]
\centering
\includegraphics[width=0.7\textwidth]{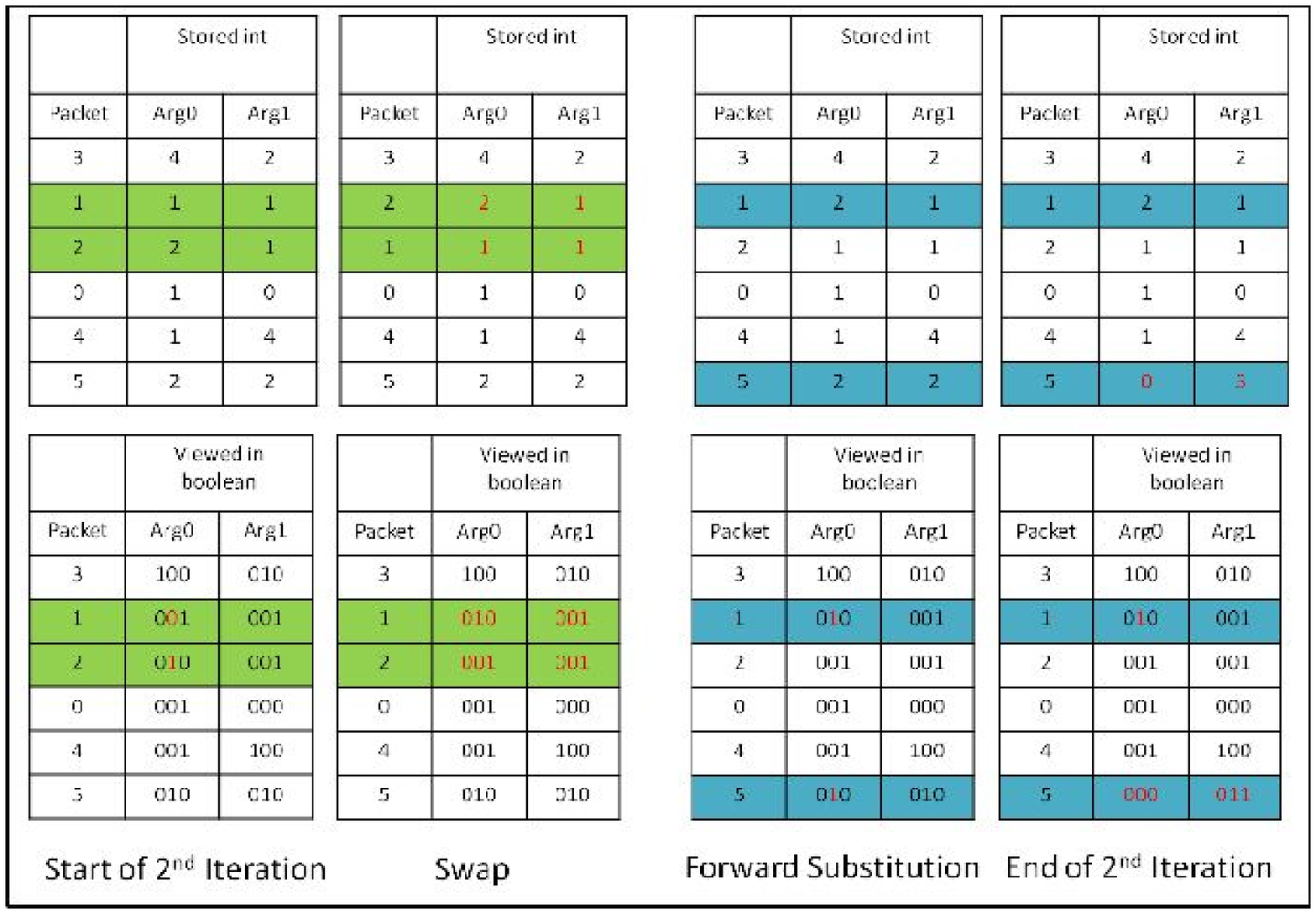}
\caption{Second step of Forward Substitution phase}
\label{fig:fs2}
\end{figure}

\item XORing phase \hfill \\
The algorithm then checks other rows to find if the first bit of the first column is also at 1. Each matching rows are then xor-ed with the first swapped row, ensuring that only the first row has the first bit positioned to 1. All data of the row are xor-ed as well.
\end{itemize}

These Swap phase and the XORing phase are repeated for each row from top to bottom until a triangle of 1 is achieved in Figure~\ref{fig:fs3}.

\begin{figure}[!hbt]
\centering
\includegraphics[width=0.25\textwidth]{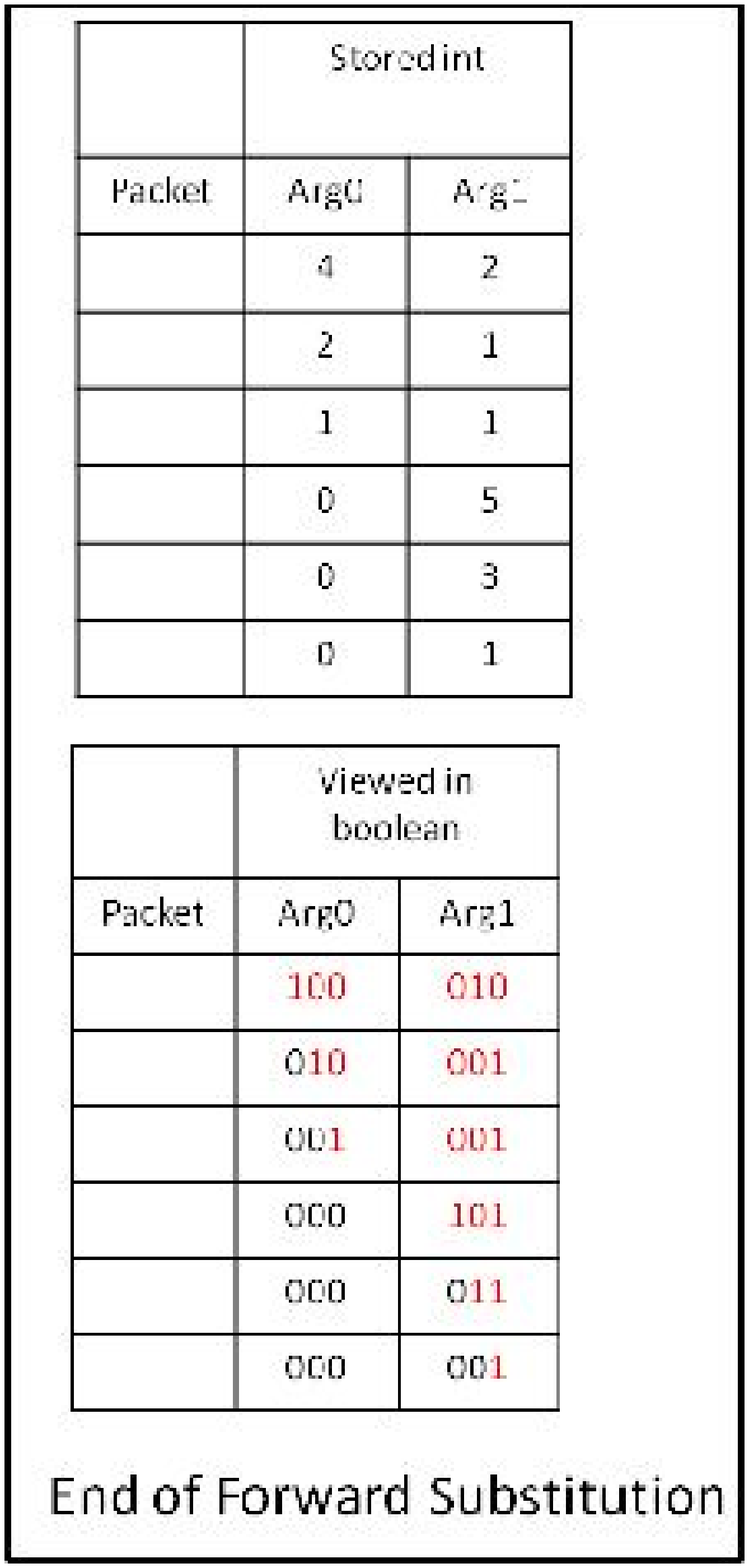}
\caption{Final state of Forward Substitution phase}
\label{fig:fs3}
\end{figure}

\item Reverse Elimination

Figures~\ref{fig:re1} to~\ref{fig:re3} show the reverse elimination process. It starts from the bottom of the matrix. From the bottom matrix last sub column, rows are scanned from bottom to up to ensure only that row is set at 1 in the sub column. Should a 1 be found, a XORing phase is applied and that row is xor-ed along with its data to remove the 1. The algorithm then proceeds on next rows till a diagonal line of 1, like in Figure~\ref{fig:re3}, is achieved.

\begin{figure}[!hbt]
\centering
\includegraphics[width=0.73\textwidth]{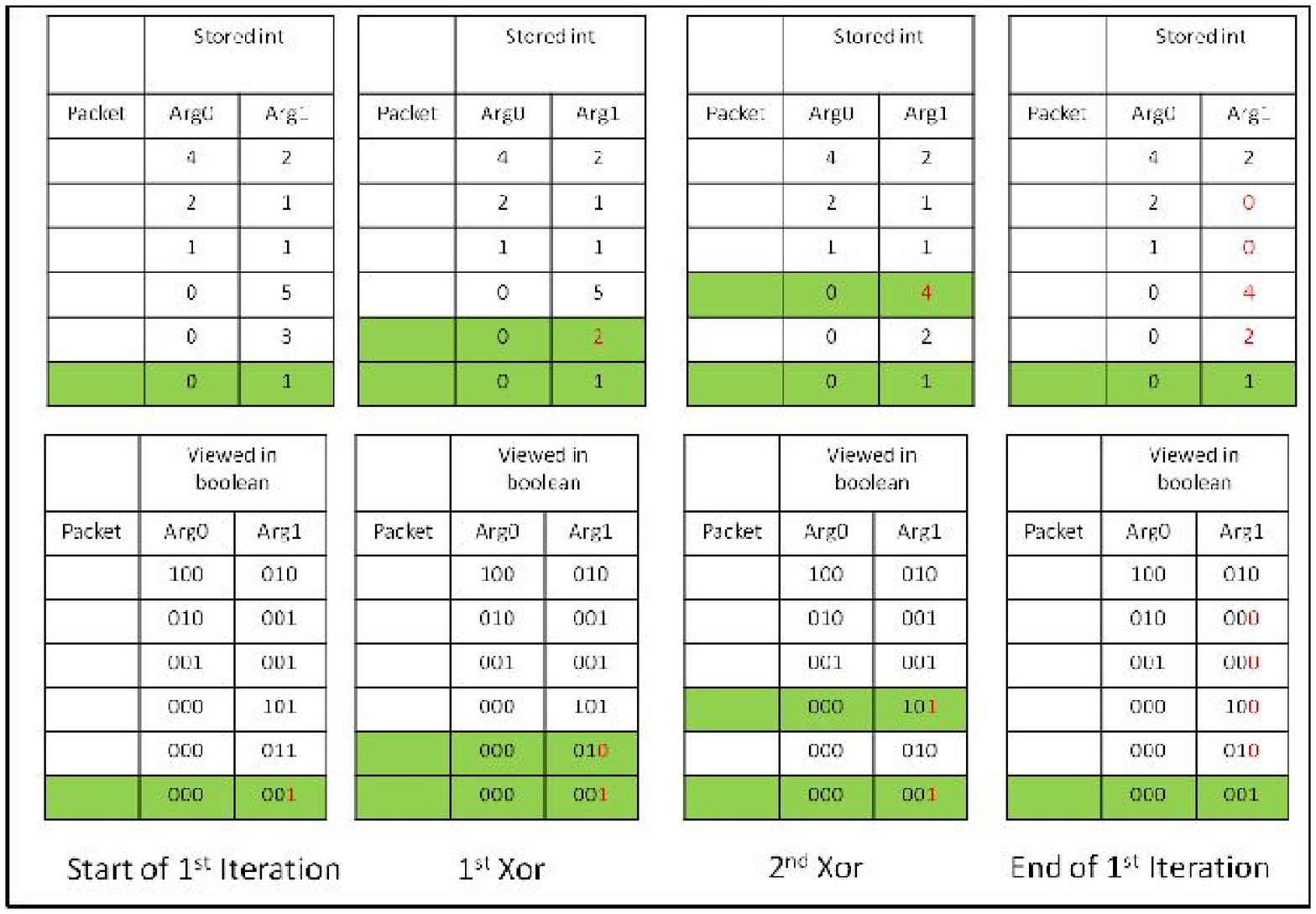}
\caption{First step of the Reverse Elimination phase}
\label{fig:re1}
\end{figure}

\begin{figure}[!hbt]
\centering
\includegraphics[width=0.73\textwidth]{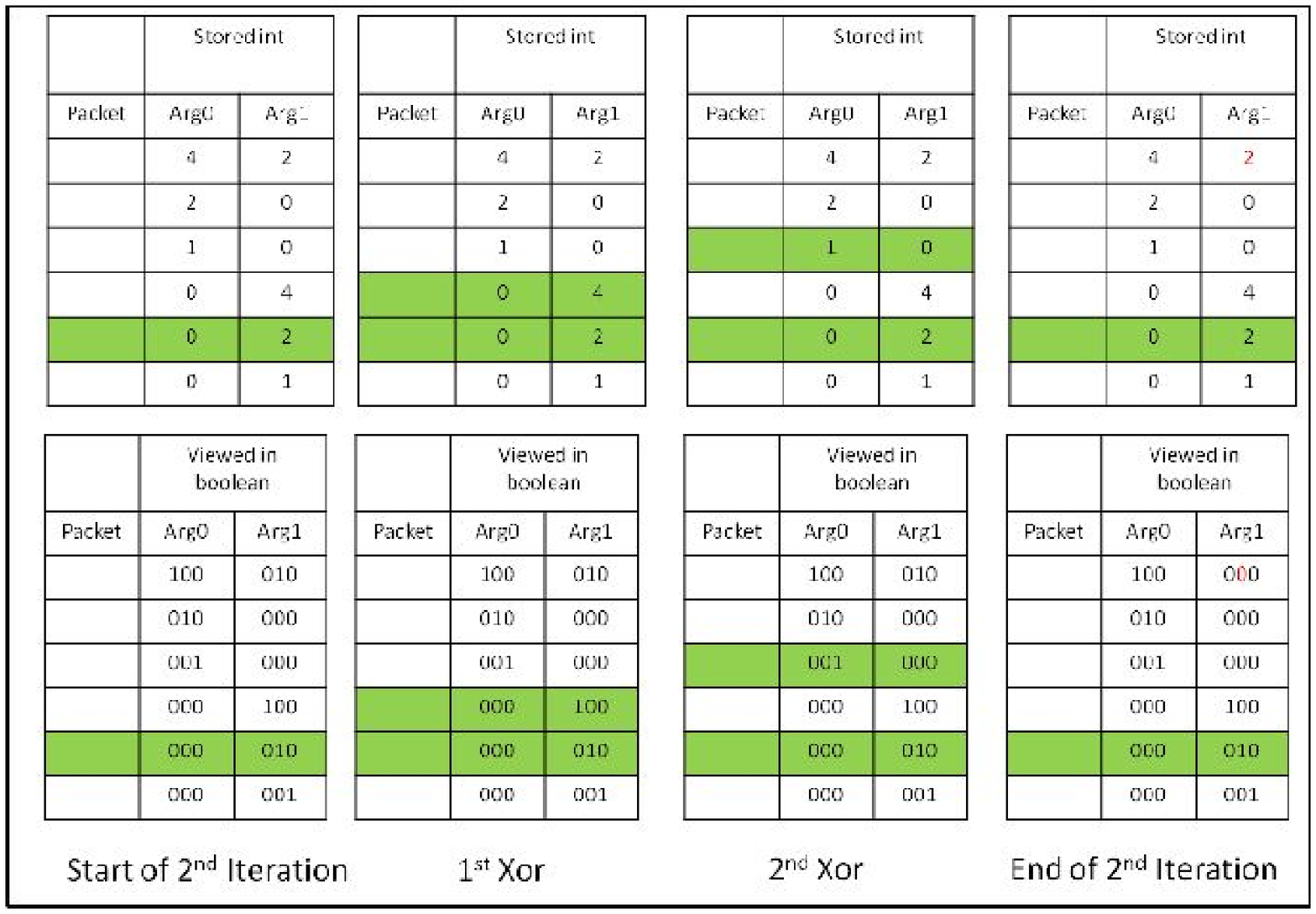}
\caption{Second step of the Reverse Elimination phase}
\label{fig:re2}
\end{figure}

\begin{figure}[!hbt]
\centering
\includegraphics[width=0.25\textwidth]{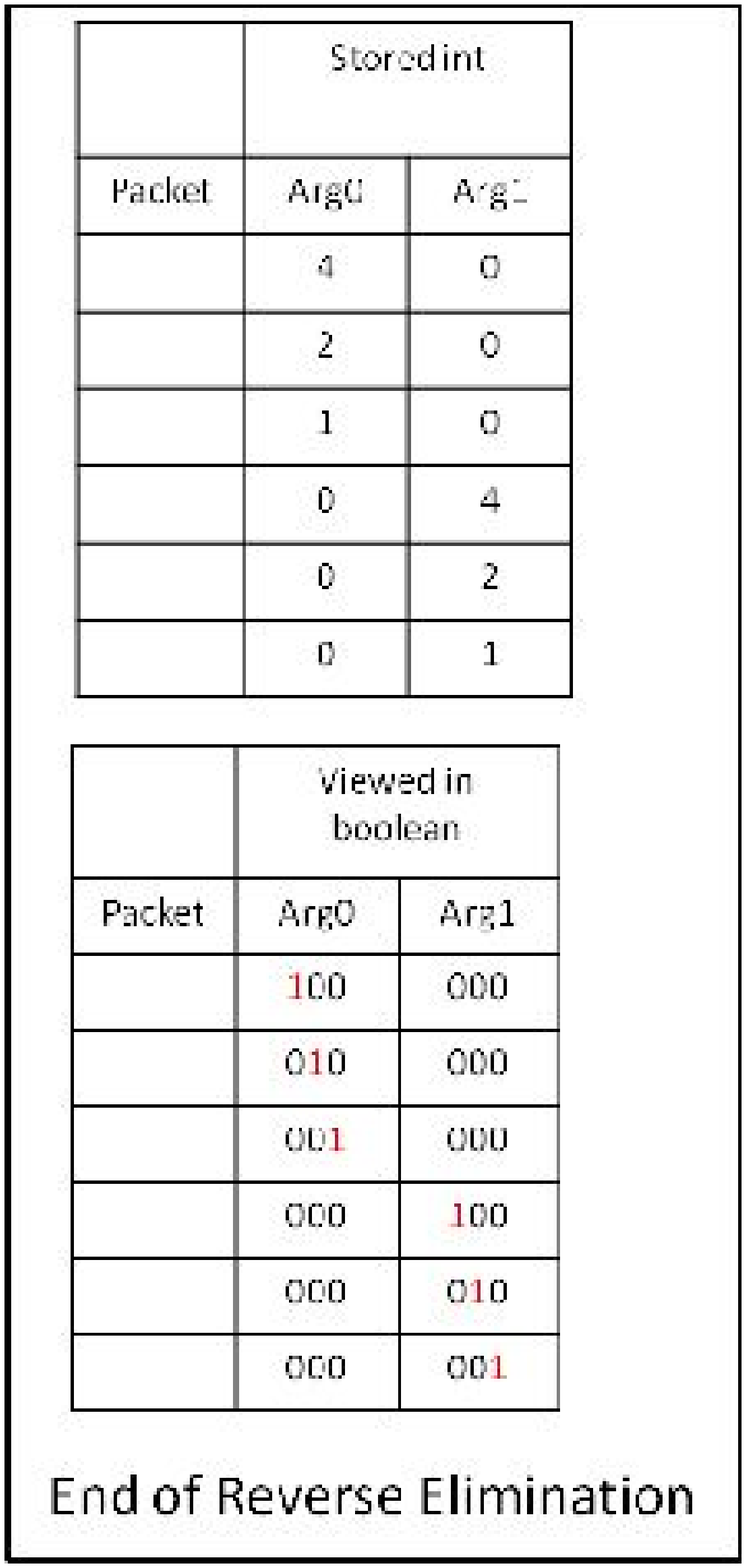}
\caption{Final state of the Reverse Elimination phase}
\label{fig:re3}
\end{figure}

\newpage

\item Reconstruction of packets

Figure~\ref{fig:rc1} and Figure~\ref{fig:rc2} show how the reconstruction is done. A reconstruction packet storage is used to separate the fragmented packets from the reconstructed ones (the original data storage may be use later). From the top, (size of mask) number of rows are xor-ed to the first packet of the reconstruction storage to reconstruct the original packet. The data is Xor-ed as well. Then the same process of (size of mask) number of rows xor-ing is applied until the end of the matrix.

\begin{figure}[!hbt]
\centering
\includegraphics[width=0.73\textwidth]{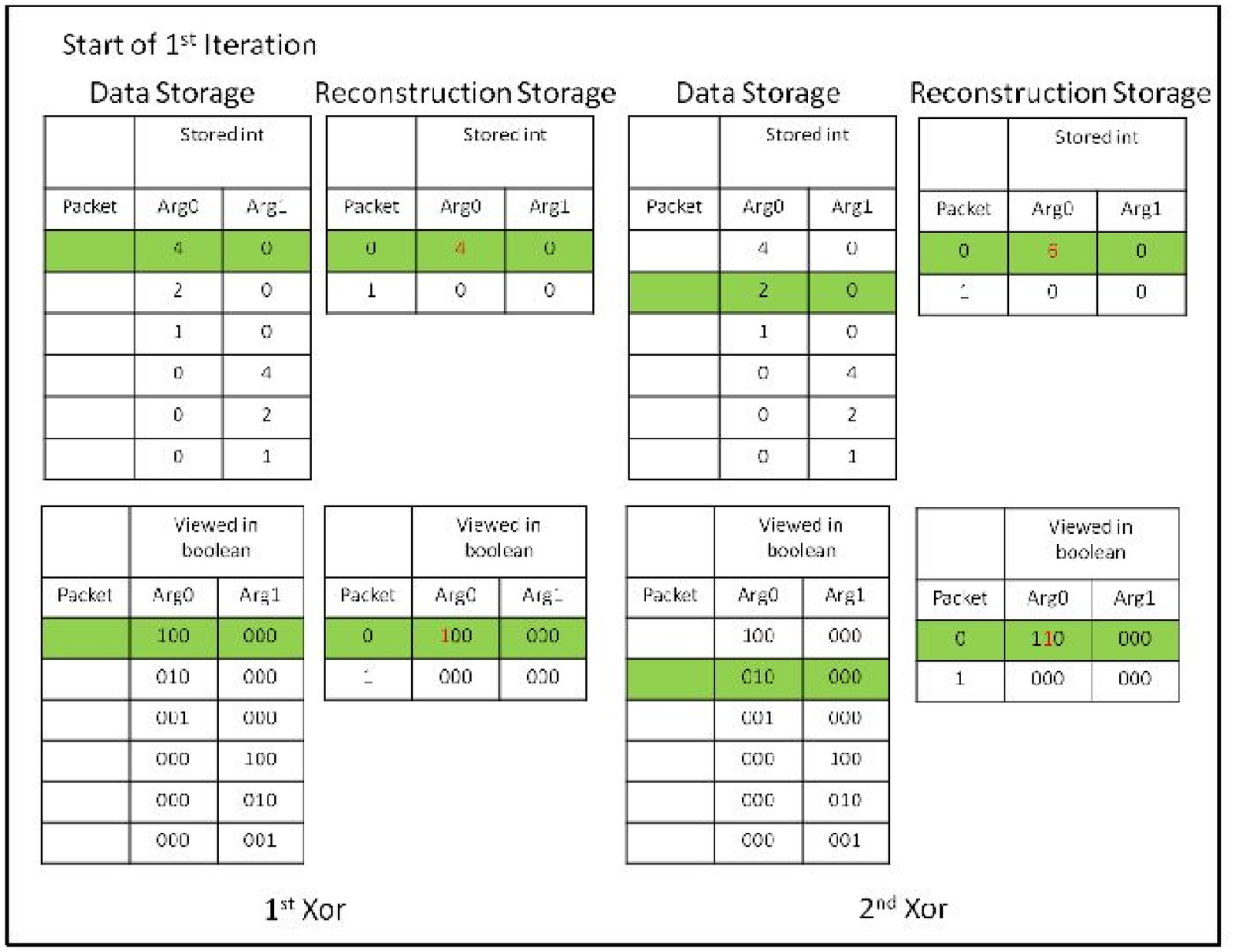}
\caption{First xor-ings of the Reconstruction phase}
\label{fig:rc1}
\end{figure}

\begin{figure}[!hbt]
\centering
\includegraphics[width=0.73\textwidth]{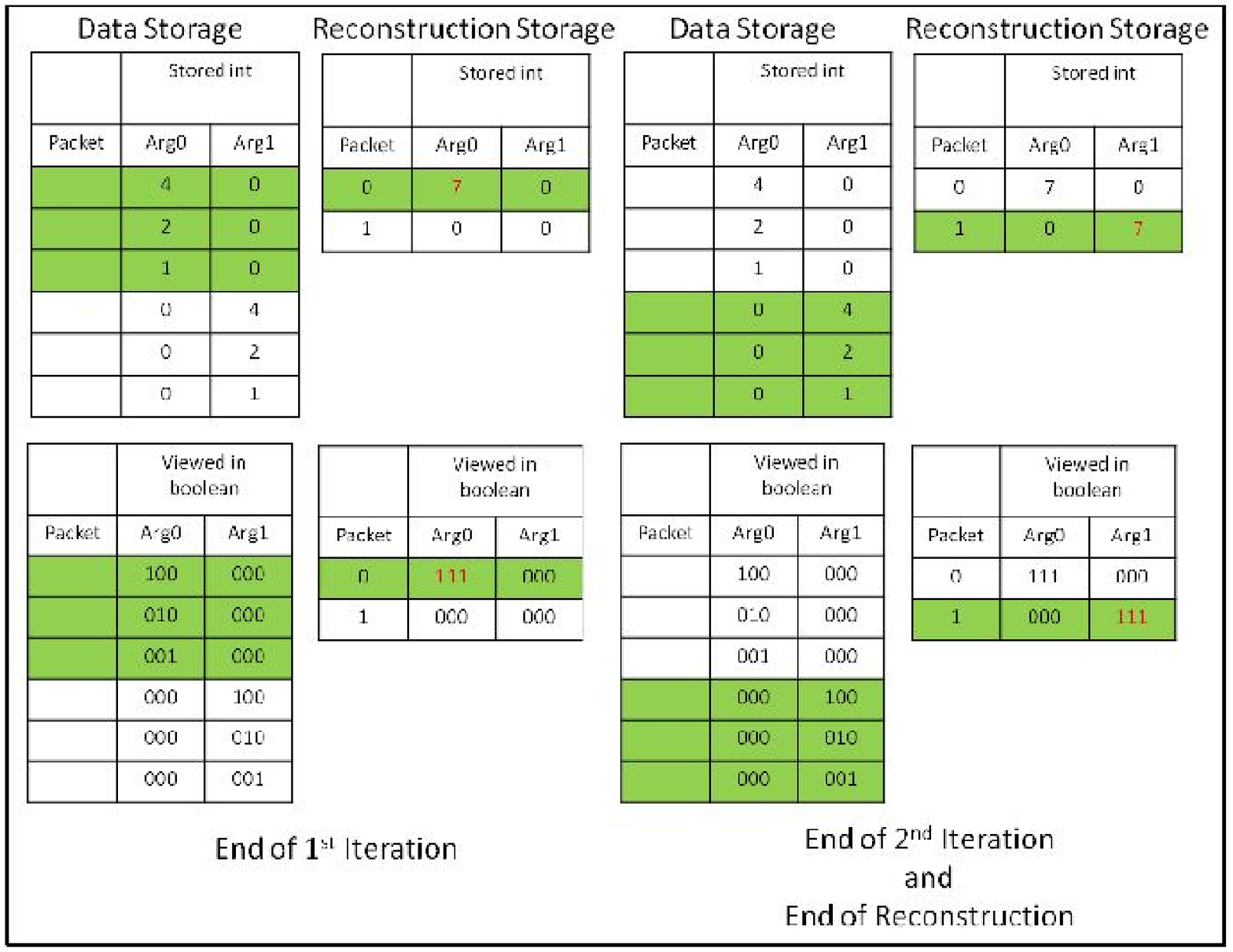}
\caption{End of xor-ings, final state of Reconstruction phase}
\label{fig:rc2}
\end{figure}

\end{enumerate}

\newpage

\subsection{Log API Definition}
\label{section:log-api}

WSNet produces ASCII prints on standard output. \texttt{DTNNC\_functions\_log.h} defines some useful functions for printing out statistics logs and debug outputs. \texttt{DTNNC\_functions\_log.h} includes node common variables outputs: id, position, etc.

\begin{lstlisting}[language=C,caption=\texttt{DTNNC\_functions\_log.h},label=list:log]
#ifndef __DTNNC_functions_log__
#define __DTNNC_functions_log__

#include "DTNNC_functions.h"

// generic function
// Use dicitonary.h and input at choice
// Check debug first
int logFunction(call_t * c, int choice);

// Check which log chosen
int logNormal(call_t * c, int choice);

// Check which print out chosen
int logDebug(call_t * c, int choice);

// Function to print out node type
int print_node_type(call_t * c);

// Function to print out node id
int print_nodeid(call_t * c);

// Function to print out node position
int print_nodePosition(call_t * c);

// Function to print out failure to print log
// Can add what to print when failed to print log
// Currenly printing node type id and position
int print_log_failed(call_t * c,int choice);

#endif // __DTNNC_functions_log__
\end{lstlisting}

\texttt{DTNNC\_log\_normal.h} outputs statistics about 'normal' events of the Network Coding module use: number of packets received/sent, encoding/decoding/ linear checking, etc.

\begin{lstlisting}[language=C,caption=\texttt{DTNNC\_log\_normal.h},label=list:logn]
#ifndef __DTNNC_log_normal__
#define __DTNNC_log_normal__

#include "DTNNC_functions_log.h"

// print send event
int print_send_event(call_t *);

// print receive event
int print_receive_event(call_t *);

// print store event
int print_store_packet(call_t *);

// print drop packet event when packet is repeat packet
int print_drop_packet_repeat(call_t *);

// print drop packet event when node has already decoded packets
int print_drop_packet_decoded(call_t *);

// print linear check fail event
int print_linear_check_failed(call_t *);

// print linear check pass event
int print_linear_check_passed(call_t *);

// print event that node has decoded packets
int print_decoded(call_t *);

// print event that node has encoded packets
int print_encoded(call_t *);

// print event of destroying or unsettting node at end of simulation
// may print information like counters and such held in node here
int print_unsetnode(call_t *);

#endif // __DTNNC_log_normal__
\end{lstlisting}

\texttt{DTNNC\_log\_debug.h} implements debugging log functions. These functions print the internal node state.

\begin{lstlisting}[language=C,caption=\texttt{DTNNC\_log\_debug.h},label=list:logd]
#ifndef __DTNNC_log_debug__
#define __DTNNC_log_debug__

#include "DTNNC_functions_log.h"

// prints what is in data storage
int print_node_storage(call_t *);

// prints what is in data holder
int print_data_holder(call_t *);

// prints what is in reconstructed packets
int print_reconstructed_packets(call_t *);

#endif // __DTNNC_log_debug__
\end{lstlisting}

Debug outputs are data-packet specific. These functions do not show only the packet header but output also the packet data to check its correctness. Therefore changing the dummy packet data structure will require to reimplement these functions. For instance, such implementation of the \texttt{DTNNC\_log\_debug.h} API needs to be adapted.
\begin{lstlisting}
printf("] dataA [%d], dataB [%d], dataC [%d], dataD [%d] \n",((struct packet_data *)getDataAt(c,i))->packetdata,((struct packet_data *)getDataAt(c,i))->packetdataB,((struct packet_data *)getDataAt(c,i))->packetdataC,((struct packet_data *)getDataAt(c,i))->packetdataD);
\end{lstlisting}

\newpage

\section{Conclusion}

This technical report has described the implementation of a \emph{Network Coding module for Wireless and Mobile DTN} in WSNet - a Wireless Sensor Network simulator. This module provides a generic framework that includes: 
\begin{itemize}
\item Programming Interfaces that defines a generic DTN node and its functionalities: IP packet storing, selecting/dropping, encoding/decoding.
\item Implementations for the main Network Coding functionalities: random selecting, random linear coding over $F_{2^n}$, Gaussian Elimination decoding.
\end{itemize}
Programming Interfaces has been generically defined to allow an easy specialization for future different coding strategies: source/destination-oriented, intra/inter-flow, application-oriented, social-oriented.

\newpage

\bibliographystyle{plain}
\bibliography{RT-0405}

\end{document}